\documentclass[aps,showpacs,prl]{revtex4}
\usepackage{amssymb,amsfonts,amsmath}

\begin{document}

\title{Alternative experimental evidence for chiral restoration in excited baryons}

\author{L. Ya. Glozman}
\affiliation{Institute for Physics, Theoretical Physics Branch, University of Graz, Universit\"atsplatz 5, A-8010 Graz, Austria}

\begin{abstract}
Given  existing empirical spectral patterns of excited hadrons it has  been
suggested that chiral symmetry is approximately restored in excited hadrons
at zero temperature/density (effective symmetry restoration). If correct, this  implies
that  mass generation mechanisms and physics in excited hadrons is very different as
compared to the lowest states. One needs an alternative and independent experimental
information to confirm  this conjecture. Using  very general chiral symmetry
arguments it is shown  that  strict
chiral restoration  in a given excited nucleon forbids its decay into the $N\pi$ channel. Hence
those excited nucleons which are assumed from the spectroscopic patterns to be in approximate 
chiral multiplets must only "weakly" decay into the $N\pi$ channel, $(f_{N^*N\pi}/f_{NN\pi})^2 \ll 1$.
However, those baryons which have no chiral partner must decay strongly with a decay
constant comparable with  $f_{NN\pi}$. Decay constants can be extracted from the
existing decay widths and branching ratios. It turnes out that for all those well established
excited nucleons which can be classified into chiral doublets $N_+(1440) - N_-(1535)$, $N_+(1710) - N_-(1650)$, 
$N_+(1720) - N_-(1700)$, $N_+(1680) - N_-(1675)$, $N_+(2220) - N_-(2250)$, $N_+(?) - N_-(2190)$,
$N_+(?) - N_-(2600)$, the ratio is  $(f_{N^*N\pi}/f_{NN\pi})^2 \sim 0.1$
or much smaller for the high-spin states. In contrast, the only 
well established excited nucleon for which the
chiral partner cannot be identified from the spectroscopic data, $N(1520)$, has  a decay constant
into the $N\pi$ channel that is comparable with  $f_{NN\pi}$. This gives an independent
experimental verification of the chiral symmetry restoration scenario. 
\end{abstract}

\maketitle

{\bf 1. Introduction.} The question of  mass generation in QCD and the
related question about interconnections of chiral symmetry and confinement
are  central  for QCD in the infrared. To answer these questions one must understand spectral
and other properties of hadrons in the light quark sector. A long history
of studies in this area has not  suggested any satisfactory answer.
 Given  existing
spectroscopic patterns it has recently been suggested that chiral and $U(1)_A$
symmetries of the QCD Lagrangian, strongly broken in the vacuum, get approximately
restored in excited hadrons \cite{G1,CG1,CG2,G2}, for a review see ref. \cite{G3}. This would imply that  physics and in
particular  mass generation mechanisms in the lowest and excited hadrons are
very different. 

The nucleon excitation spectrum is shown in Fig. 1 of ref. \cite{G3}. Only 
well-established states have to be seriously considered. It is well seen that there is no
obvious chiral partner to the nucleon. This implies that  chiral symmetry breaking 
effects are very strong in the given case and consequently the chiral symmetry is realized 
nonlinearly in the nucleon \cite{WEIN}. This is  consistent with the well-known result
that  the mass of the nucleon is (at least mostly) induced by the quark condensate
of the vacuum \cite{IOFFE}. This  is  supported by the  empirical
success of the Goldberger-Treiman relation, which also implies that the
nucleon mass comes  from the chiral symmetry breaking in the vacuum  \cite{LEE} .
Another clear indication of the strong chiral symmetry
breaking effects in  the nucleon is a large pion-nucleon coupling constant. All
empirical successes of  chiral perturbation theory for  the nucleon \cite{M}
also implicitly rely on the assumption that the nucleon mass is due to
chiral symmetry breaking in the vacuum. These results, taken together,
suggest that there is no chiral partner to the nucleon.

Obvious approximate parity doublets are observed in the 1.7 GeV region. These have
been assigned to the $(0,1/2)+(1/2,0)$ representation of the parity-chiral
group because there are no approximately degenerate doublets in the same mass
region in the spectrum of the delta-resonance \cite{CG2,G3}.
 A clear testable prediction of
the chiral symmetry restoration scenario is  an existence of  chiral partners
 of the well established high-lying resonances $N(2190)$ and $N(2600)$. A dedicated
 experimental search of these missing states can be undertaken \cite{S}.
A similar situation takes place in the Delta-spectrum.

While these parity doubling patterns are impressive, they are still only suggestive,
because so far no other complementary experimental data would independently
tell us that these parity doublets are due to effective chiral symmetry
restoration. Strict chiral restoration in a given baryon would imply that its
diagonal axial charge is zero and hence the diagonal coupling constant to the pion
must vanish \cite{G3,GN,JPS1,JPS2}. An experimental verification of the smallness of these
quantities for approximate  parity doublets would be a confirmation of the chiral
restoration scenario. However, it is unclear how to extract this information
from the existing data. Hence it is important to find alternative experimental
signatures of chiral restoration.

A different assignement of excited baryons  was discussed in ref. \cite{JHK}, where some of the
excited nucleons below the 1.7 GeV region have been combined with the delta-resonances
into quartets of the $(1,1/2)+(1/2,1)$ representation. Such an
assignement requires  a large pion-baryon coupling constant  \cite{JHK},
 because there is a large mass splitting between the assumed members
 of the quartets, and hence does not
correspond to the effective chiral symmetry restoration regime.

There is  rich experimental data on strong decays of excited hadrons. It
turnes out that the effective chiral restoration  implies a very strong
selection rule. Namely, it predicts that if chiral symmetry is completely
restored in a given excited nucleon ($B$), then it cannot decay into the $\pi N$ channel,
 i.e. the coupling constant $f_{BN\pi}$ must vanish. This selection rule is based
exclusively on general properties of chiral symmetry, what will be demonstrated
below, and hence is model-independent. Such a coupling can be
extracted from the existing decay data \cite{CK,RB}. Then an independent verification
of the chiral symmetry restoration scenario would require that those excited
nucleons that are approximate parity doublets, only "weakly" decay into the
$\pi N$ channel,  $(f_{B N\pi}/f_{NN\pi})^2 \ll 1$, and, on the contrary,
those excited nucleons, to which the chiral partner cannot be identified
from the spectroscopic data, must decay strongly, 
$(f_{B N\pi}/ f_{NN\pi})^2 \sim 1$.

\medskip
{\bf 2.} Let us review first the question whether there are or not
chiral symmetry constraints for a magnitude of the $\pi NN$ coupling
(well known from the sigma-model derivations \cite{GL,WEIN,LEE}). 
As a starting point for the chiral symmetry analysis it is always useful to consider the linear 
realization of chiral symmetry. The standard
assumption is that the pion  field  (together with the sigma field or quark condensate)
transform as $(1/2,1/2)$ of $SU(2)_L \times SU(2)_R$. Throughout the paper we
ignore small quark masses. The very absence of an independent chiral partner to the nucleon 
implies that the nucleon field  $N$ in the Wigner-Weyl mode transforms as
$(0,1/2)+(1/2,0)$ and is massless. Its chiral partner is $\gamma_5 N$.
The axial transformation law is then

\begin{equation}
N \rightarrow 
\exp \left(  \imath  \gamma_5 \frac{\theta^a_A\tau^a}{2} 
\right)N.
\label{nax}
\end{equation}

\noindent
 Then one can construct the  $\pi N N$ vertex as

\begin{equation}
\sim \bar N e^{i \gamma_5 \frac{{\vec \pi} \cdot {\vec \tau}}{f_\pi}} N.
\label{lsm}
\end{equation}

\noindent
This vertex is  chiral-invariant (chiral scalar) because the axial rotation
of the nucleon field is compensated by the rotation of the pion field. 
A strength of the interaction is fixed by the coupling constant which is an
external parameter and can have any arbitrary value.
Hence we conclude
that chiral symmetry does not restrict the $\pi NN$ interaction which
can be arbitrarily strong. In the Nambu-Goldstone mode the nucleon
field becomes massive due to its coupling with the chiral order parameter.
The axial current conservation connects this nucleon mass and axial charge
via the Goldberger-Treiman relation with the pion-nucleon coupling constant,
$g_{NN\pi} = \frac{g_A M_N}{f_\pi}$.
Hence the large pion-nucleon coupling constant encodes a physical origin
of  the nucleon mass as due to chiral symmetry breaking in the vacuum.
In other words, the large $\pi NN$ coupling constant can be used as a natural
unit for strong chiral symmetry breaking effects in a baryon. 

\medskip
{\bf 3.} Assume that we have a free $I=1/2$ chiral doublet $B$ in the $(0,1/2)+(1/2,0)$
representation  and there are no chiral symmetry breaking terms.  This
doublet is a column \cite{LEE}

\begin{equation}
B = \left(\begin{array}{c}
B_+\\
B_-
\end{array} \right),
\label{doub}
\end{equation}

\noindent
where the bispinors
$B_+$ and $B_-$ have positive and negative parity, respectively,
because the parity  on the doublet space is defined to be

\begin{equation}
P:~~~~ B(\vec x, t) = \sigma_3 \gamma_0 B(-\vec x, t).
\end{equation}

\noindent
The chiral transformation law
 under the $(0,1/2) \oplus (1/2,0)$ representation 
 provides a mixing of two fields  $B_+$ and $B_-$ \footnote{Note that
 the axial transformation given in \cite{LEE} is incorrect as it
 breaks chiral symmetry of the kinetic term. The correct axial
 transformation is given in ref. \cite{G3}.}

\begin{equation}
B \rightarrow 
\exp \left( \imath \frac{\theta^a_V \tau^a}{2}\right)B; ~~
B \rightarrow 
\exp \left(  \imath \frac{\theta^a_A\tau^a}{2} \sigma_1
\right)B.
\label{VAD}
\end{equation}

\noindent
Here $\sigma_i$ is a Pauli matrix that acts in the  $2 \times 2$
space of the parity doublet. 
Then  the chiral-invariant Lagrangian of the free parity doublet is given as

\begin{eqnarray}
\mathcal{L}_0 & = & i \bar{B} \gamma^\mu \partial_\mu B - m_0 \bar{B}
B  \nonumber \\
& = &  i \bar{B}_+ \gamma^\mu \partial_\mu B_+ + 
i \bar{B}_- \gamma^\mu \partial_\mu B_-
- m_0 \bar{B}_+ B_+ - m_0 \bar{B}_- B_- .
\label{lag}
\end{eqnarray}

\noindent
Alternative forms for this Lagrangian can be found in refs. \cite{TK,TIT}.

A crucial
property of this Lagrangian  is that  the fermions
$B_+$ and $B_-$ are exactly degenerate and
have a nonzero chiral-invariant mass $m_0$. In contrast, for
usual  (Dirac) fermions chiral symmetry in the Wigner-Weyl mode 
restricts particles to be massless.

From the axial transformation law (\ref{VAD}) one can read off the
axial charge matrix, which is $\gamma_5 \sigma_1$. Hence the diagonal axial
charges of the opposite parity baryons are exactly 0, $g_+^A = g_-^A = 0$,
while the off-diagonal axial charge is 1,  $ |g_{+-}^A| = |g_{-+}^A| = 1$.
This is another crucial property that distinguishes the parity
doublets from the Dirac fermions where $g^A = 1$.
The axial vector current conservation translates this axial charge matrix
into the $\pi B_{\pm}B_{\pm}$ coupling constants which are zero. Hence a small
(vanishing) value of the pion-baryon coupling constant taken together
with the large baryon mass would tell us that the origin of this mass is not
due to chiral symmetry breaking in the vacuum. 
As discussed in the introduction
part an experimental verification of the smallness of the diagonal axial
charges or smallness of the pion-baryon coupling constants would be a direct
verification of the chiral symmetry restoration scenario in excited nucleons. It is 
unclear, however, how to measure these quantities.

\medskip
{\bf 4.} Consider a possible $B_\pm N \pi$ vertex.
Since these parity doublet baryons $B_+, B_-$ are in the
$(0,1/2) \oplus (1/2,0)$ representation, like the nucleon,
on the first value the chiral-invariant vertex
$ B N \pi$ is possible. This is incorrect, however. The reason is
that the axial transformation laws for the nucleon field and for the
parity doublet fields are very different, because they live in different
linear spaces. While the axial rotation of the nucleon
field is compensated by the chiral transformation of the pion field
in the vertex (\ref{lsm}), so this vertex is chiral-invariant, such a
compensation is impossible if one of the legs of this vertex is
substituted by $B_\pm$. Consequently one cannot construct a chiral-invariant
interaction vertex $B_{\pm}N\pi$. 
 Hence decay of the parity doublets with
completely restored chiral symmetry into $N \pi$ is 
impossible.\footnote{This statement can also be proven in the following
way. Assume that a $\pi N$ decay of an exact parity doublet
is possible. Then there must be a self-energy contribution
$B_\pm \rightarrow \pi N \rightarrow B_\pm$ into its mass. Then
the axial rotation  (\ref {VAD}) would require that the S-wave
$\pi N$ state transforms into the P-wave $\pi N$ state. However,
in the Nambu-Goldstone mode the axial rotations of the pion and
nucleon states are fixed - these are the nonlinear axial transformations
\cite{JPS1,JPS2}. Given these well known axial transformation
properties of the Goldstone boson and nucleon it is not possible
to rotate the S-wave
$\pi N$ state  into the P-wave $\pi N$ state. Therefore, there cannot be
any $\pi N$ self-energy component in $B_\pm$. 
Hence a decay
$B_\pm \rightarrow \pi N$ is forbidden.}
However,
their decay into e.g. $N\rho$ or $N\pi\pi$ is not forbidden.

If, in contrast, the excited baryon has no chiral partner, then its
mass, like in the nucleon case is exclusively due to chiral symmetry breaking
in the vacuum. Its axial charge should be comparable with the nucleon
axial charge. Then  nothing forbids its strong decay into $N\pi$. One then
expects that the decay coupling constant should be of the same order
as the pion-nucleon coupling constant. These two extreme cases suggest
that a magnitude of the $BN\pi$ decay constant can be used as an indicator
of the mass origin. 

\medskip
{\bf 5.}
In reality, of course, chiral symmetry is never completely restored in excited
hadrons. To account a small amount of chiral symmetry breaking contribution
one must add into Lagrangian terms  that provide a coupling of the given hadron with
the vacuum. Then the baryon mass can be split into two different parts: 
(i) chiral-invariant $m_0$ and (ii) chiral "non-invariant"  $m'$. The latter is
due to coupling of the hadron with the chiral-noninvariant vacuum 
(i.e. with the quark condensate).  However, if the coupling with the quark condensate 
is weak and
consequently the chiral symmetry breaking mass $m'$
 is small, $m' \ll m_0$, then the decay into $N\pi$ must be strongly suppressed.
 The magnitude of $m'$ can be determined from the
chiral asymmetry parameter \cite{G2,G3}

\begin{equation}
\chi = \frac{|M_{B_+} - M_{B_-}|}{ M_{B_+} + M_{B_-} }.
\label{ass}
\end{equation}

\noindent
This papameter can be interpreted as a part of the hadron mass due to
chiral symmetry breaking in the vacuum.
 Chiral asymmetries
of the excited nucleons in the 1.7 GeV mass region and above are typically
within 0.02. Hence, if these parity doublets are indeed due to the effective symmetry restoration,
then their decay  into $N \pi$ must be very strongly suppressed.

\medskip
{\bf 6.}
Now we can formulate predictions of the chiral symmetry restoration scenario.
If a state is a member of an approximate chiral multiplet and $\chi \ll 1$, then its
decay into $N \pi$ must be strongly suppressed, 
$(f_{B N\pi}/f_{NN\pi})^2 \ll 1$. If, on the contrary, this excited hadron
has no chiral partner and hence its mass is due to chiral symmetry breaking in the vacuum,
then it should strongly decay into $N \pi$ and hence  
$(f_{B N\pi}/ f_{NN\pi})^2 \sim 1$. In the following section we demonstrate
that this prediction is in accord with the existing data.

\medskip
{\bf 7.}  
Given the well-known Rarita-Schwinger formalism for the higher-spin fields,
one can construct  phenomenological $BN\pi$ Lagrangians. Then decay constants
 $f_{BN\pi}$ can be extracted from the $B \rightarrow N + \pi$ decay widths,
 see e.g. \cite{CK,RB}. The pion-nucleon coupling constant is well-known,
 $f_{NN\pi} =1.0$. In Table 1 we show ratios $(f_{BN\pi}/f_{NN\pi})^2$ for
 all well-established states. It is well seen that this ratio is $\sim 0.1$
 or smaller for approximate $J=1/2,3/2,5/2$ parity doublets. For the high-spin
 states this ratio is practically vanishing. This is consistent with the
 recent demonstration of the large J-rate of chiral restoration within the only 
 known exactly solvable
 confining and chirally-symmetric model \cite{WG}.
 In contrast, the only well-known
 state $J=3/2, N_-(1520)$, where a chiral partner is missing in the spectrum,
 decays very strongly into the $N\pi$ channel. Hence chiral symmetry breaking
 effects are as large in this case as in the nucleon. Consequently its
 physical origin should be very different as compared to all other resonances
 which can be classified into parity doublets.
 
 In conclusion, the effective chiral symmetry restoration conjecture is now
 supported by strong decays of excited nucleons. A decay of the strict parity
 doublets into the $N\pi$ channel is forbidden by chiral symmetry, assuming that
 there is no independent chiral partner to the nucleon. Hence a fraction $\Gamma_i/\Gamma$
 of the $N\pi$ decay channel  becomes small, even though decays into other
 channels (not protected by chiral symmetry) are strongly  suppressed by the
 phase space factor.

\acknowledgments
L.Ya.G. is grateful to Tom Cohen for useful comments and to Dan Riska for
correspondence. Support of the Austrian Science Fund through grant
P19168-N16 is  acknowledged.
\begin{table}
\begin{center}
\caption{Chiral multiplets of excited nucleons.
Comments: (i) All these states are well established and
can be found in the Baryon Summary Table of the Review of Particle 
Physics.  (ii) There
are two possibilities to assign the chiral representation:
$(1/2,0) \oplus (0,1/2)$ or $(1/2,1) \oplus (1,1/2)$ because
there is a possible chiral pair in the $\Delta$ spectrum
with the same spin with similar mass. (iii) The missing  chiral partner
is predicted.
 }
\begin{tabular}{|llllll|} \hline
Spin & Chiral multiplet &  Representation & $\chi$ & $(f_{B_+N\pi}/f_{NN\pi})^2 -  (f_{B_-N\pi}/f_{NN\pi})^2$ & Comment\\ \hline
1/2& $N_+(1440 ) - N_-(1535)$ & $(1/2,0) \oplus (0,1/2)$ &
 0.032    &  0.15 - 0.026   & (i) \\

1/2& $N_+(1710) - N_-(1650)$ & $(1/2,0) \oplus (0,1/2)$ &
0.02   & 0.0030 - 0.026  & (i) \\

3/2& $N_+(1720) - N_-(1700)$ & $(1/2,0) \oplus (0,1/2)$ &
0.01   & 0.023 - 0.13    & (i) \\

5/2&$N_+(1680) - N_-(1675)$ & $(1/2,0) \oplus (0,1/2)$ &
0.002   & 0.18 - 0.012  & (i) \\

7/2&$N_+(?) - N_-(2190)$ &  see comment (ii)  &
?   &  ? - 0.00053  & (i),(ii),(iii) \\

9/2&$N_+(2220) - N_-(2250)$ &
 see comment (ii) &
0.01   &  0.000022 - 0.0000020 & (i),(ii) \\

11/2&$N_+(?) - N_-(2600)$ &   see comment (ii) &
?   &  ? - 0.000000064  & (i),(ii),(iii) \\

\hline
\hline
3/2& $ N_-(1520)$ & no chiral partner &
-   &   2.5   &  (i)  \\
\hline

\end{tabular}
\end{center}
\label{t3}
\end{table}


\begin{thebibliography}{99}

\bibitem{G1} L. Ya. Glozman, Phys. Lett. B {\bf  475}, 329 (2000). 
\bibitem{CG1} T. D. Cohen and L. Ya. Glozman, Phys. Rev. D {\bf 65}, 016006 (2002).
\bibitem{CG2} T. D. Cohen and L. Ya. Glozman, Int. J. Mod. Phys. A {\bf  17}, 1327 (2002).
\bibitem{G2} L. Ya. Glozman, Phys. Lett. B {\bf 539}, 257 (2002);
Phys. Lett. B {\bf 541}, 115 (2002); Phys. Lett. B {\bf 587}, 69 (2004);
Int. J. Mod. Phys. A {\bf 21}, 475 (2006).
\bibitem{G3} L. Ya. Glozman,  Phys. Rep. {\bf 444}, 1 (2007).
\bibitem{GL} M. Gell-Mann, M. Levy, Nuovo Cimento {\bf 16}, 705 (1960).
\bibitem{WEIN} S. Weinberg, Phys. Rev.  {\bf 166}, 1568 (1968).
\bibitem{IOFFE} B. L. Ioffe, Nucl. Phys. {\bf B 188}, 317 (1981); E: {\bf B 191}, 591 (1981).
\bibitem{LEE} B. W. Lee, Chiral Dynamics, Gordon and Breach, New York, 1972
\bibitem{M} U-G. Meissner, Chiral QCD: Baryon Dynamics, in: At the frontiers of Particle
Physics. Handbook of QCD., edited by M. Shifman, v. 1, p. 417, World Sc. 2001.
\bibitem{S} A. Sibirtsev, J. Haidenbauer, S. Krewald, T.-S.H. Lee, U.-G- Meissner, A. Thomas,
arXiv: 0706.0183 [nucl-th].
\bibitem{GN} L. Ya. Glozman, A. V. Nefediev, Phys. Rev.  {\bf D 73}, 074018 (2006).
\bibitem{JPS1} R. L. Jaffe, D. Pirjol, A. Scardicchio, Phys. Rev. {\bf D 74}, 057901 (2006);
Phys. Rev. Lett. {\bf 96}, 121601 (2006).
\bibitem{JPS2} R. L. Jaffe, D. Pirjol, A. Scardicchio, Phys. Rep. {\bf  435}, 157 (2006).
\bibitem{JHK} D. Jido, T. Hatsuda, T. Kunihiro, Phys. Rev. Lett. {\bf 84}, 3252 (2000). 
\bibitem{CK} W. K. Cheng, C. W. Kim, Phys. Rev. {\bf 154}, 1525 (1967).
\bibitem{RB} D. O. Riska and G. E. Brown, Nucl. Phys. {\bf A 679}, 577 (2001).
\bibitem{TK} C. DeTar, T. Kunihiro, Phys. Rev. {\bf D 39}, 2805 (1989).
\bibitem{TIT} D. Jido, M. Oka, A. Hosaka, Progr. Theor. Phys. {\bf 106}, 873 (2001).
\bibitem{WG} R. F. Wagenbrunn, L. Ya. Glozman, Phys. Lett. {\bf B 643}, 98 (2006);
Phys. Rev. {\bf D 75}, 036007 (2007).




\end{thebibliography}
\end{document}